\documentclass[runningheads]{llncs}
\usepackage{algorithm}
\usepackage{algorithmicx}
\usepackage[noend]{algpseudocode}

\usepackage{subcaption}
\usepackage{placeins}
\usepackage{comment}
\usepackage[hidelinks]{hyperref}
\usepackage{array}
\usepackage{mathtools}
\usepackage{amsmath}
\usepackage{multirow}
\usepackage{dsfont}
\usepackage{tabularx}
\usepackage{xcolor}
\usepackage{colortbl}
\usepackage{comment}
\usepackage{enumitem}

\authorrunning{Chernikova et al.}

\definecolor{blond}{rgb}{1.0, 0.95, 0.85}

\algnewcommand{\LineComment}[1]{\State \(\triangleright\) #1}

\title{Cyber Network Resilience against Self-Propagating Malware Attacks}
\author{Alesia Chernikova\inst{1}\and Nicolò Gozzi\inst{2}\and Simona Boboila\inst{1}\and Priyanka Angadi\inst{4}\and \\ John Loughner\inst{4}\and Matthew Wilden\inst{4}\and Nicola Perra\inst{3}\and Tina Eliassi-Rad\inst{1}\and Alina Oprea\inst{1}}

\institute{Northeastern University, Boston, MA, USA\and University of Greenwich, UK\and School of Mathematical Sciences, Queen Mary University of London, UK\and PricewaterhouseCoopers LLP, USA}
\begin{document}

\maketitle
\begin{abstract}
Self-propagating malware (SPM) has led to huge financial losses, major data breaches, and widespread service disruptions in recent years. In this paper, we explore the problem of developing cyber resilient systems capable of mitigating the spread of SPM attacks. We begin with an in-depth study of a well-known self-propagating malware, WannaCry, and present a compartmental model called SIIDR that accurately captures the behavior observed in real-world attack traces. Next, we investigate ten cyber defense techniques, including existing edge and node hardening strategies, as well as newly developed methods based on reconfiguring network communication (NodeSplit) and isolating communities. We evaluate all defense strategies in detail using six real-world communication graphs collected from a large retail network and compare their performance across a wide range of attacks and network topologies. We show that several of these defenses are able to efficiently reduce the spread of SPM attacks modeled with SIIDR. For instance, given a strong attack that infects 97\% of nodes when no defense is employed, strategically securing a small number of nodes (0.08\%) reduces the infection footprint in one of the networks down to 1\%.
\end{abstract}

\section{Introduction}
Self-propagating malware (SPM) has become one of the top cyber threats in recent years. In 2016, Mirai~\cite{mirai_2016} malware infected more than 600K consumer devices and launched a widespread DDoS attack targeting over 175K websites. The WannaCry~\cite{wannacry_2017} ransomware attack of 2017 affected more than 300K vulnerable devices in 150 countries in a few days, from entire healthcare systems to banks and national telecommunications companies.  Worryingly, there have been reports of its re-appearance during the COVID-19 pandemic. Recently, ransomware attacks have increased significantly, with the emergence of new threats like Ryuk (2019)~\cite{Coresecurity2022}, PureLocker (2020)~\cite{Seals2019}, and many others.
SPM campaigns attempt to exploit vulnerabilities on specific ports by blending in with legitimate traffic. Since blocking ports entirely is often not feasible, defending against SPM is particularly challenging. Machine learning (ML) techniques have been employed~\cite{PORTFILER}, with the goal of detecting the attack and taking reactive measures after the data breach has occurred. However, the performance of ML methods often degrades when there is a high number of false positives, which are hard to triage by human experts.

In this paper, we take a graph robustness perspective for \emph{proactively protecting cyber networks against self-propagating malware attacks}. We study the problem of how to build cyber resilient systems and how to configure communication in cyber networks to prevent the spread of SPM attacks. Towards this ambitious goal, our first task is understanding and modeling the  behavior of a well-known SPM malware, WannaCry, by using compartmental models that stem from epidemiology. We then turn our attention to methods for enhancing network resilience against these attacks. We model the topology of the network via the communication flows collected from real network traces, which we obtained from an industry partner. Our main insight is to analyze communication networks through the lens of graph robustness, an area that has been studied extensively in other applications (e.g., ~social-, ~information-, ~transportation-, and mobility-networks), but much less in cyber security. 

On the SPM attack modeling side, we show that prototypical models for virus propagation (such as SIS and SIR~\cite{Brauer2008}) do not fit the behavior of WannaCry well. We thus introduce a new model (SIIDR) that captures the behavior of self-propagating malware more accurately.
In particular, our new model introduces a dormant state, in which the malware is installed in the system, but is not active for some interval (this is a common behavior observed in SPM, as well as advanced persistent threat  attacks, which have been documented to be ``slow-and-low'' for months and sometimes years). We use  real traffic logs generated by multiple variants of WannaCry to  select the best fit model and estimate the parameters (i.e., transition rates) that best characterize WannaCry attacks. 

On the defense analysis side, we perform an in-depth evaluation on several complementary topological-based defenses. We investigate a large number of defense techniques (10) based on various cybersecurity strategies such as node hardening, edge hardening, isolation, and reconfiguring communication. We evaluate their ability to increase network resilience to cyber attacks using two robustness metrics that have been shown to be accurate indicators of graph connectivity, and, implicitly, of the network's resilience to attack propagation~\cite{freitas2022graph}: (a) the spectral radius of a graph and (b) the relative size of the largest connected component in the graph. We also propose two new defensive methods: \emph{NodeSplit}, which reconfigures the nodes with the largest number of incident edges by migrating half of their communication links to new nodes; and \emph{Community Isolation}, which constructs communities in the graph and then strategically hardens edges that connect the communities to thwart the attack propagation. Hybrid strategies that combine NodeSplit with edge hardening (e.g., setting up firewall rules) are particularly successful at minimizing the spectral radius of a graph (our first robustness metric), achieving over 60\% reduction on most of our graphs, with a small budget of only 50 split nodes and a fraction of 0.1 secured edges. Hybrid NodeSplit + Community Isolation strategies perform very well on the second robustness metric, being able to break down the largest connected component to less than 20\% of its original size on most of the studied graphs. 

To evaluate our defenses in realistic conditions, we use six real-world communication flow  graphs collected from a large retail network. These graphs model application communication on well-known ports (22, 80, 139, 383, 443, 445) and are up to 620K nodes and 6.8 million edges. We thoroughly evaluate and compare the defense strategies in terms of their ability to reduce the spread of malware modeled with SIIDR and the budget required for slowing down the attack.
Node hardening techniques are the most effective defenses over a wide range of attack scenarios and network topologies studied here, leading to a $20\times$ decrease in the effective attack strength. This reduction results in substantial infection footprint minimization. For instance, given a strong attack that infects 97\% of nodes when no defense is employed, strategically securing a small number of nodes (i.e., 50 nodes, which account for 0.08\% of the nodes on port 22) reduces the infection footprint on port 22 down to only 1\%.

We summarize our contributions below:
\begin{itemize}[itemsep=0em]
    \item We propose and evaluate SIIDR, a compartmental attack model that captures the behavior of SPM accurately. We use real SPM traffic logs from  WannaCry to estimate the attack parameters. 
    \item We perform an in-depth evaluation of 10 defense techniques and compare them using two graph robustness metrics. 
    \item We introduce two novel defenses: NodeSplit (to reconfigure communication of top-degree nodes) and Community Isolation (to  harden edges between communities), and show their effectiveness particularly in hybrid strategies. 
    \item We evaluate the effectiveness of various defense strategies against SPM attacks, using six large real-world communication graphs. 
    \item We provide recommendations on the effectiveness and cost of multiple defenses to inform network operators on various proactive defense options against SPM attacks. Our open-source code is available on GitHub~\cite{our_codebase}.
\end{itemize}

\section{Problem Statement and Background}
In a recent survey, Freitas et al.~\cite{freitas2022graph} note that the study of graph vulnerability and robustness is still nascent in cybersecurity. Existing research includes modeling lateral attack movement between computers and analytical studies of interdependent spatial networks~\cite{chen2019,chencascadian2018,freitasd2m2020}. However, they point out the need for additional work in several directions, including comprehensive evaluations of various attack and defense scenarios. Our work directly addresses this need, and is, in that sense, particularly timely.  
We study both facets of building cyber resilient systems, attacks and defenses, in an integrated and complementary way.

In cybersecurity, eliminating or mitigating vulnerabilities is achieved through ``system hardening''~\cite{Zlotnik2021} and depending on where or how protection is applied, it can refer to network hardening, server hardening, operating system hardening, application hardening, etc. Reducing the ``attack surface'' consists in addressing known vulnerabilities via changing passwords, removing unused services, applying security patches, closing open network ports, configuring firewalls and setting intrusion-detection systems. On the theoretical side, most of the previous techniques for increasing network robustness are based on classic mathematical epidemiology results, which link the spreading of a virus of a graph with its spectral radius~\cite{Chakrabarti2008,Prakash2011}. The epidemics dies out quickly if the spectral radius is less than a threshold, which depends on the virus propagation method. Hence, topological changes are employed to bring the spectral radius below this threshold. 

\vspace{3pt}
\noindent \textbf{Threat model:}
We aim to design efficient defense strategies that increase network robustness against SPM attacks. We consider cyber networks such as enterprise networks, data center networks, or cloud systems, in which communication flows between nodes can be modeled as a graph. 
We assume that the malware first compromises one victim machine on the network (``patient zero''), after which it spreads to other vulnerable machines within the network over a specific protocol, e.g., HTTP (port 80), SSH (port 22), SMB(port 445), etc. 

We derive a realistic attack model (namely, SIIDR) and its parameters by running an actual WannaCry attack, under homogeneous mixing assumptions~\cite{vespignani2012}. Homogeneous mixing models imply that all hosts have identical rates of infection-spreading contacts. Our attack experiments were carried out within a local subnet, where such an assumption is valid. For WannaCry modeling, we analyze Zeek logs collected at the border of the monitored subnet. We assume that these logs are not compromised by the attacker.  

To design and evaluate defense strategies, we use communication data of a large retail network. We assume that the network traffic has not been compromised, and, thus, the logged connections can be used to derive an accurate graph representation of the network communication topology.

\vspace{3pt}
\noindent \textbf{Challenges:}
Building cyber-resilient systems is challenging for multiple reasons. First, realistic modeling of actual attacks is difficult, due to the limited availability of attack traces, and the ethical considerations that prevent us from recreating known attacks in real-world networks. Second, the continual evolution of attacks that attempt to avoid detection requires innovative proactive measures that are able to counter a wide range of potential threats.
Third, building resilient infrastructures is budget-dependent both in terms of infrastructure and software updates, as well as human effort. Careful assessments of complete eradication strategies versus mitigation (containment) methods are necessary to establish real-world feasibility of the defenses.

\section{SPM Modeling}
\label{sec:spm_modeling}
Infectious disease research has inspired the study of mathematical models for malware propagation, in part due to the similarities in the behavior of human viruses and computer viruses. Many of these epidemiological models are compartmental. That is, the population is divided into states (a.k.a.~classes), based on the health status of individuals and the type of disease~\cite{Brauer2008}. Examples include susceptible ($S$), exposed ($E$), infectious ($I$), quarantined ($Q$), and recovered ($R$), to name a few. 
In this study, we are investigating the specific case of self-propagating malware. All types of self-propagating malware have one defining characteristic: once a host becomes infected, it starts probing other computers on the Internet randomly, with the goal of spreading the infection widely. This type of behavior guides the mathematical modeling of SPM. 

We use real-world attack traces to model self propagating malware, and derive compartmental models and parameters that closely fit actual attack propagation data.   Model fitting for deriving best models and their parameters from data has been widely used in the study of infectious diseases~\cite{liao2015,stocks2020}, but less so in the modeling of computer viruses. To the best of our knowledge, we are the first to model self-propagating malware based on real attack traces to find the model and parameters that most accurately describe a real attack.
In the remainder of this section we describe the data, selection methodology and results that identify the model that captures SPM behavior most precisely.

\vspace{3pt}
\noindent \textbf{WannaCry data: } We select WannaCry malware as a representative self-propagating malware attack, which can be configured with multiple parameters to generate a range of propagation behaviors. As shown by Ongun et al.~\cite{PORTFILER}, other SPM malware such as Mirai, Hajime, and Kenjiro follow similar propagation patterns, and our attack modeling will likely generalize to other attacks.
We set up a virtual environment featuring the EternalBlue Windows exploit that was used in the 2017 WannaCry  attack. External traffic is blocked in order to ensure isolation of the virtual environment. Initially, one of the virtual machines is infected with WannaCry, and then the attack starts spreading, as infected IPs begin to scan other IPs on the same network. We identified two characteristics that WannaCry uses to control its spread: 1) the number of threads used for scanning, and 2) the time interval between scans. We conducted multiple experiments by running WannaCry with different characteristics and collecting log traces with Zeek network monitoring tool.

\vspace{3pt}
\noindent \textbf{Epidemics reconstruction: }
We use the WannaCry traces to study the malware behaviour and reconstruct the epidemics. 
The start and end time of the epidemics for each WannaCry variant is given by the first malicious attempt, and the last communication event, respectively. Hence, the first IP trying to establish a malicious connection represents ``patient zero'', and an IP trying to establish at time $t$ a malicious connection for the first time is considered infected at time $t$. Based on the WannaCry traces, we make the following observations:

\begin{itemize}
    \item Distribution of $\Delta t$ intervals between attacks: We observe that the $\Delta t$ between two consecutive attacks from the same infected IP is not fixed. This heterogeneous distribution of $\Delta t$ intervals between subsequent attacks from the same infected IP introduces the idea of a $I \leftrightarrow I_D$ dynamics, where $I$ represents the infectious state and $I_D$ represents the dormant state.
    \item Distribution of $\Delta t$ intervals between last attack and end of trace: The $\Delta t$ between the last attack from an infected IP and the end of observations is quite large. The non-zero, high-valued and heterogeneous distribution of $\Delta t$  between last attack from an infected IP and the end of observation time supports the idea of a $I \rightarrow R$ dynamics, with $R$ being the recovered (previously infected, but not infectious anymore) state.
\end{itemize}

\begin{figure}[t]
\centering
\includegraphics[width=0.35\textwidth]{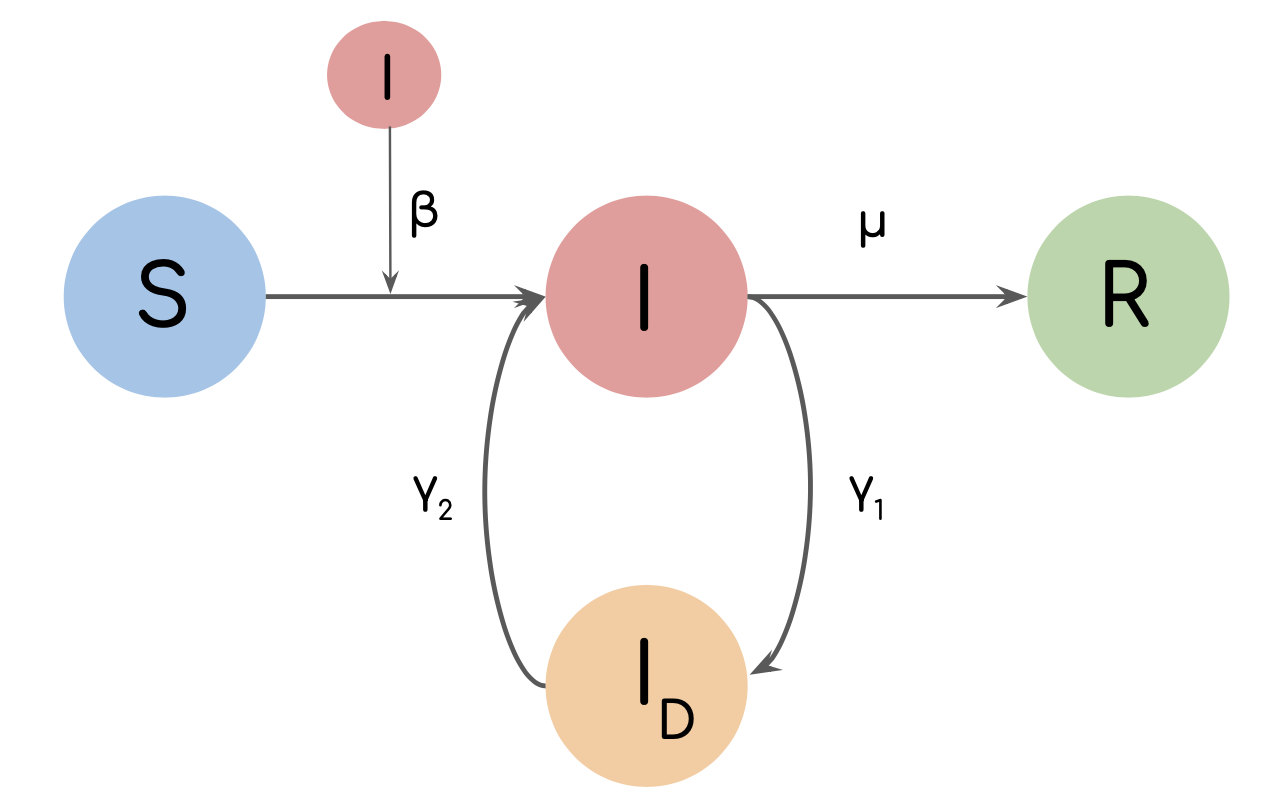}
\vspace{-10pt}
\caption{Compartmental structure and transitions of the SIIDR model. Susceptible nodes (S), exposed to infected IPs, acquire the infection with rate $\beta$. Infected nodes transition with rate $\gamma_1$ or recover with rate $\mu$. From infected dormant state ($I_D$) nodes transition back to $I$ with rate $\gamma_2$.}
\label{fig:SIIDRstates}
\vspace{-15pt}
\end{figure}

\noindent \textbf{Model description: }
In accordance with the behavior observed from WannaCry traces, we propose SIIDR, an extension to the SIR model~\cite{Brauer2008} that includes the infected dormant state $I_D$. 
While other models like SEIR or SEIRS have been used to model the spread of malware~\cite{rey2015}, SIIDR more closely explains the behavior observed of WannaCry. Furthermore, a model like SEIRS would not make sense, given that Recovery means patching of the EternalBlue vulnerability. Once the operating system is updated (patched), it will not become susceptible again to the same vulnerability.
Figure ~\ref{fig:SIIDRstates} presents the transition diagram corresponding to the SIIDR model.
A node that is infected may either recover with rate $\mu$, or move to the dormant state with rate $\gamma_1$. 
From the dormant state, it may become actively infectious again with rate $\gamma_2$. 
We assume a homogeneous mixing model~\cite{vespignani2012}, that is, every node is potentially in contact with all the others. This is a good approximation because the WannaCry attack experiments were run within a subnet, where every node was able to scan every other internal IP within the same subnet. The system of differential equations that describe the dynamics of SIIDR, the derivation of the basic reproduction number of the model and discussion of stability are presented in a companion paper~\cite{Chernikova_SPM}.

\bgroup
\setlength\tabcolsep{3pt}
\begin{table*}[t] 
\centering
\begin{tabular}[t]{|c||c|c|c|c|}
\hline
\textbf{WannaCry} & \multirow{2}{*}{\bf SI} &\multirow{2}{*}{\bf SIS} &\multirow{2}{*}{\bf SIR} &\multirow{2}{*}{\bf SIIDR}\\
\textbf{variant} & &&&\\
\hline
wc\_1\_500ms & 583& 143& 114& \textbf{-126}\\
\hline
wc\_1\_1s &431 &188 & 145&\textbf{-127}\\
\hline
wc\_1\_5s& 683&163& 143& \textbf{72}\\
\hline
wc\_1\_10s & 462&197 &53&\textbf{-92}\\
\hline
wc\_1\_20s &704 &\textbf{559} &696 &700 \\
\hline
wc\_4\_500ms &277&76&-45 &\textbf{-166}\\
\hline
wc\_4\_1s &222&160 & 107&\textbf{-55}\\
\hline
wc\_4\_5s &412&186&158&\textbf{-46}\\
\hline
\end{tabular}
\quad
\begin{tabular}[t]{|c||c|c|c|c|}
\hline
\textbf{WannaCry} & \multirow{2}{*}{\bf SI} &\multirow{2}{*}{\bf SIS} &\multirow{2}{*}{\bf SIR} &\multirow{2}{*}{\bf SIIDR}\\
\textbf{variant} & &&&\\
\hline
wc\_4\_10s & 513&94&-36&\textbf{-145}\\
\hline
wc\_4\_20s &606 &76&11 &\textbf{-117}\\
\hline
wc\_8\_500ms& 375&101&18&\textbf{-147}\\
\hline
wc\_8\_1s &178&91& 51&\textbf{-116}\\
\hline
wc\_8\_5s &149&104&-35&\textbf{-121}\\
\hline
wc\_8\_10s &253 &74&-90&\textbf{-118}\\
\hline
wc\_8\_20s &387 &164&173&\textbf{-89}\\
\hline
\end{tabular}
\vspace{3pt}
\caption{AIC scores for each of the SPM models for different WannaCry variants. Each WannaCry variant is identified by two parameters: the number of threads used for scanning and the time interval between scans (i.e., wc\_1\_500ms uses 1 thread to scan every 500 ms).}
\label{tab:aic_scores}
\vspace{-10pt}
\end{table*}
\egroup

\vspace{3pt}
\noindent \textbf{Model selection and parameter estimation: }
To determine which model is the best fit, we compare common epidemiological models SI, SIS, SIR with SIIDR using the Akaike information criterion (AIC)~\cite{akaike1973information}. The best fit is the model with the minimum AIC score. The AIC scores for all candidate models are presented in Table~\ref{tab:aic_scores}.
We observe that the SIIDR model has the lowest AIC scores overall, except for wc\_1\_20s (on which all models have high AIC). For instance, the AIC score  for the SIIDR model for variant wc\_4\_500ms is as low as -166, while the AIC scores for SI, SIS, and SIR models are 277, 76, and -45, respectively.   Hence, SIIDR fits the WannaCry dynamics better than the other candidate models.

We use Sequential Monte Carlo (SMC)~\cite{mckinley2018approximate,MINTER2019100368} to approximate the posterior distribution of rates $(\beta,\mu,\gamma_1,\gamma_2)$ that fit the actual WannaCry data.  The WannaCry variants with less than $ 20\%$ infected IPs (i.e., 7 infections) were excluded due to insufficient samples to generate accurate models. Table~\ref{tab:siidrparams} lists the mean values of SIIDR parameters posterior distributions. $\texttt{dt}$ is the simulation time step. We have one contact per $\texttt{dt}$, thus, the transmission probability over a contact-link equals $\beta$. We use these parameters to evaluate defenses in Section~\ref{sec:siidr_sim}. For more details on model selection and parameter estimation, see our companion paper~\cite{Chernikova_SPM}.
\begin{table*}[!t]
\centering
\begin{tabular}[t]{|c||c|c|c|c||c|}
\hline
\textbf{WannaCry} &\multirow{2}{*}{$\beta$}
&\multirow{2}{*}{$\mu$} &\multirow{2}{*}{$\gamma_1$} &\multirow{2}{*}{$\gamma_2$}
&\multirow{2}{*}{$\texttt{dt}$}\\
\textbf{variant} & &&&&\\
\hline
wc\_1\_500ms &0.10 &0.06 &0.76 &0.04 &0.09\\
\hline
wc\_1\_1s & 0.11 &0.07 &0.71 &0.07 &0.06\\
\hline
wc\_1\_5s &0.37 & 0.52&0.27 & 0.44&0.16\\
\hline
wc\_1\_10s &0.12 & 0.06&0.75 & 0.05&0.09\\
\hline
wc\_4\_1s &0.14 & 0.07&0.75 & 0.08&0.05\\
\hline
wc\_4\_5s &0.12 & 0.07& 0.76& 0.07&0.07\\
\hline
wc\_8\_20s &0.13 & 0.09& 0.74& 0.08& 0.07\\
\hline
\end{tabular}
\vspace{3pt}
\caption{Mean values from posterior distribution of SIIDR parameters, estimated with sequential Monte Carlo. For details, see our companion paper~\cite{Chernikova_SPM}.}
\vspace{-20pt}
\label{tab:siidrparams}
\end{table*}

\section{Defense Methodology}
We model the network as a host-to-host communication graph. The nodes represent systems like computers, mainframe, peripherals, load balancer devices, etc. that communicate over TCP/UDP. An edge exists between two systems if they exchange network communication. We create these communication graphs from NetFlow data collected inside the organization. 
%We thoroughly investigate several defenses intended to secure the network against cyber attacks by reducing the ``attack surface''. 
Putting defensive control on some of the nodes and edges in the network makes them inaccessible to the attacker. We define the \texttt{Attacker's Reachability Graph} (ARG) as the nodes and edges that an attacker has access to after defenses have been applied.   
We devise and study defense strategies from four different perspectives, as illustrated in Figure~\ref{fig:strategies} and summarized in Table~\ref{tab:defenses} We describe these defenses next:
\begin{figure*}[!t]
\centering
  \begin{subfigure}[l]{0.24\linewidth}
    \includegraphics[width=\linewidth]{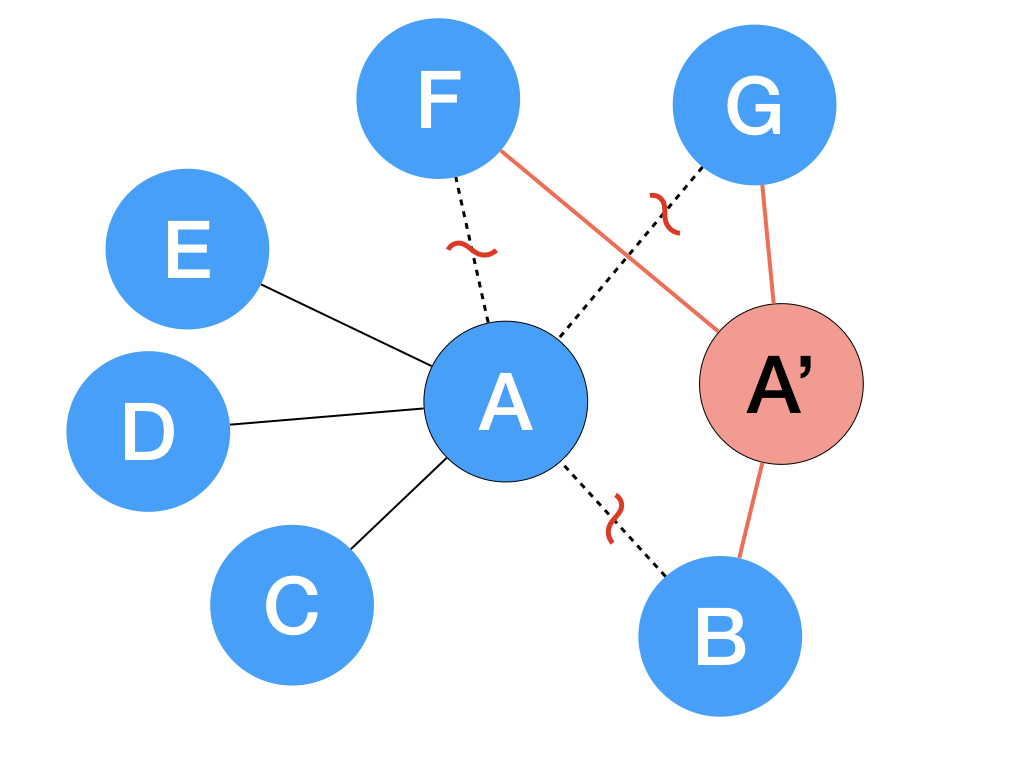}
    \caption{Node\\Splitting}
  \end{subfigure}
   \begin{subfigure}[l]{0.24\linewidth}
    \includegraphics[width=\linewidth]{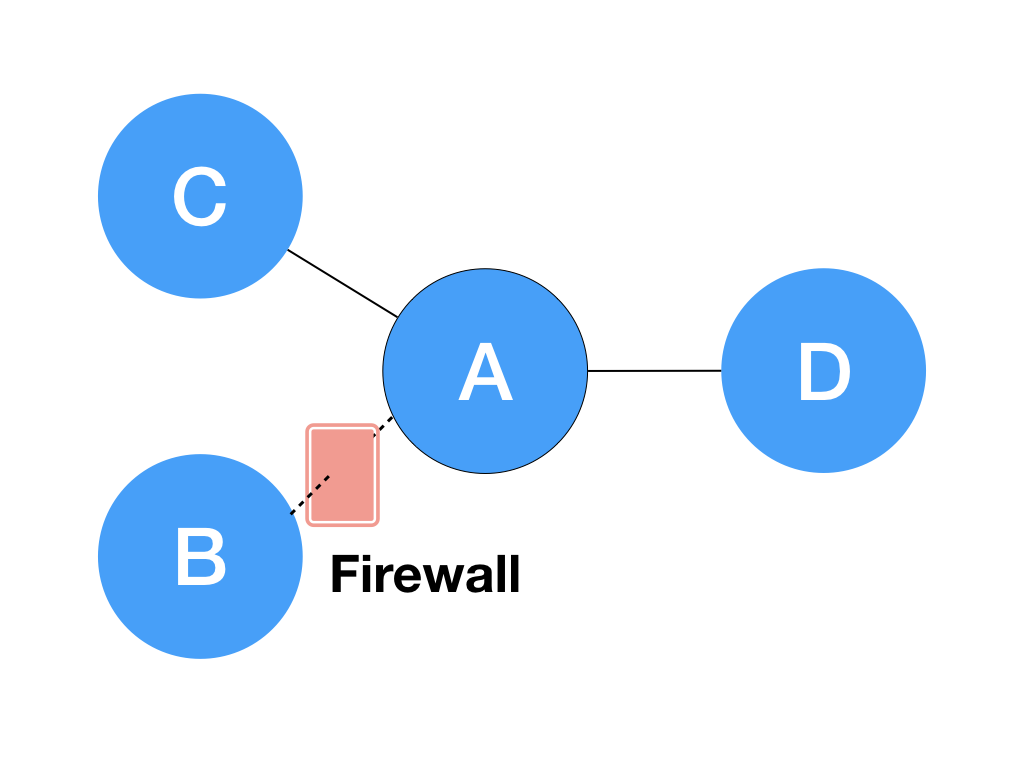}
   \caption{Edge\\Hardening}
  \end{subfigure}
  \begin{subfigure}[l]{0.24\linewidth}
    \includegraphics[width=\linewidth]{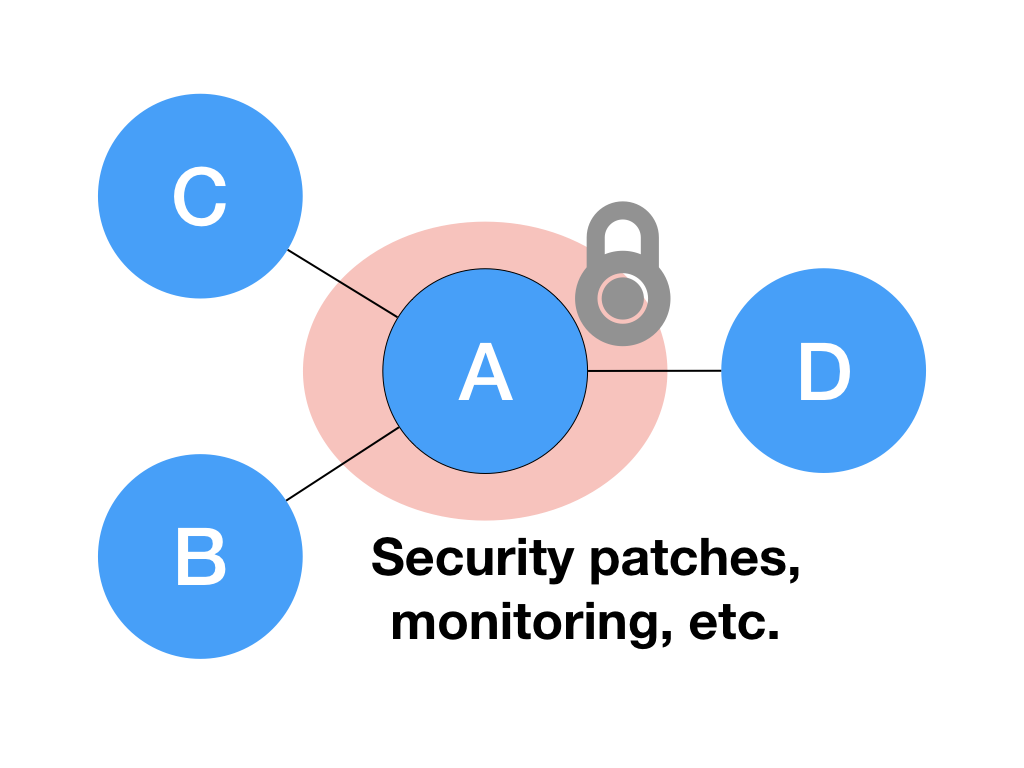}
   \caption{Node\\Hardening}
  \end{subfigure}
    \begin{subfigure}[l]{0.24\linewidth}
    \includegraphics[width=\linewidth]{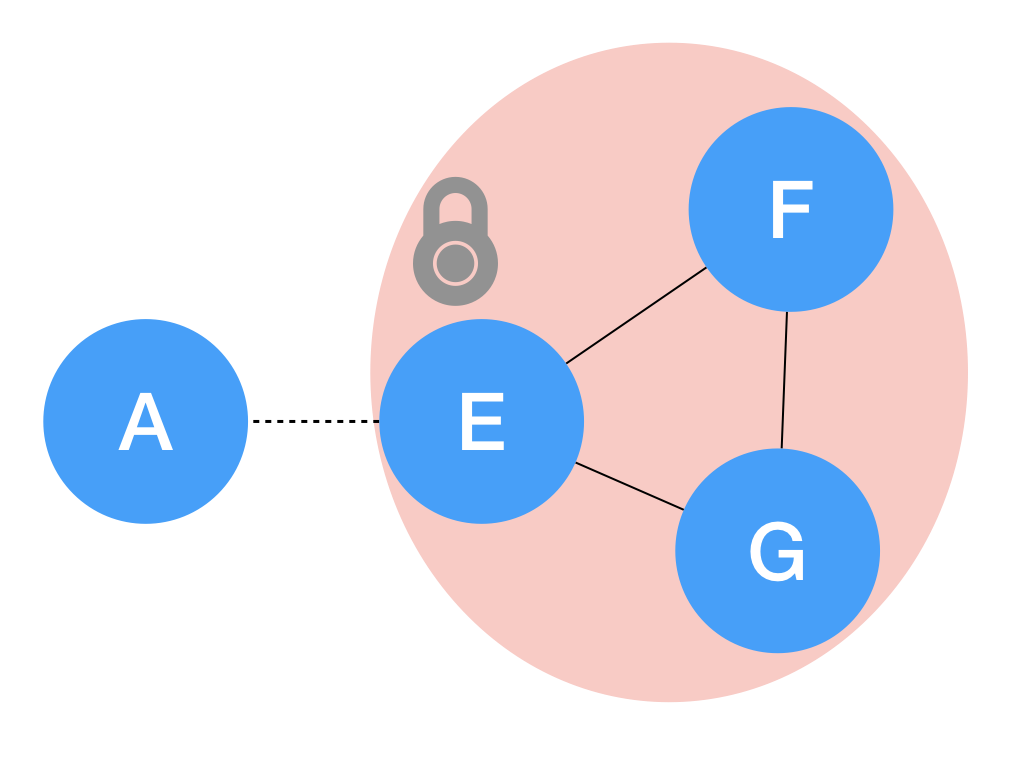}
   \caption{Isolation}
  \end{subfigure}
\caption{Cybersecurity defense strategies: a) Node splitting -- access to nodes B, F, G from node A is disabled, and a new node A' is created, with links to nodes B, F, G. b) Edge hardening via firewall rules. c) Node hardening via security patches and monitoring techniques. d) Isolation -- the denser cluster comprised of nodes E, F, G is separated from the rest of the network through enhanced security measured at the boundary.}
\vspace{-10pt}
\label{fig:strategies}
\end{figure*}

\begin{table*}[!b]
\centering
\begin{tabular}{|>{\centering\arraybackslash}m{1in} |>{\centering\arraybackslash}m{1.2in} |>{\centering\arraybackslash}m{1.2in} 
|>{\centering\arraybackslash}m{1.2in} |}
\hline 
Defense strategy & Algorithms &Topological changes in ARG & Cybersecurity measures \\
\hline\hline
Node splitting &NodeSplit &Node addition and edge rewiring &Reconfiguring access rules  \\ \hline
Edge hardening &MET~\cite{le_2015}, RandE  & Edge removal &Firewalls, closing open ports\\\hline
Node hardening &Degree, ENS, NB~\cite{torres_2021}, RandN  & Node removal & System updates, Security patches\\\hline
Isolation &CI-Edge, CI-Node  & Edge / node removal & Hardening at community boundary\\
\hline
\end{tabular}
\caption{Defense strategies  and associated topological changes.}
\label{tab:defenses}
\vspace{-25pt}
\end{table*}

\begin{itemize}

\item \textbf{Node Splitting:} This newly proposed method reconfigures network communication by splitting and balancing top-degree nodes. It requires node addition and rewiring of edges, which  can be implemented by reconfiguring communication via traffic management solutions available in data centers.

\item \textbf{Edge Hardening:} This implies monitoring certain edges in the communication graph and corresponds to edge removal from ARG. It can be implemented via firewall rules for blocking certain flows, closing of some  network ports,  or adding rules to network intrusion detection systems (NIDS).   

\item \textbf{Node Hardening:} This implies monitoring certain nodes in the communication graph and  corresponds to node removal from ARG. It can be implemented via system updates, security patch installation, and  monitoring through endpoint agents.

\item  \textbf{Isolation:} This newly introduced method requires hardening of edges or nodes on community boundaries  via separation of densely connected communities from the rest of the network. 
\end{itemize}

% Table~\ref{tab:defenses} summarizes the defense strategies analyzed in this study, their algorithmic implementations, associated topological changes, as well as possible cybersecurity measures used to implement these defenses. 

\vspace{1pt}
\noindent \textbf{Node Splitting: }
We introduce NodeSplit, a new algorithm that reconfigures network communication in order to increase resilience to SPM attacks.
In our proposed approach, node splitting is targeted at top-degree nodes. If compromised, these highly connected nodes become super-spreaders and potentially propagate the infection within a large portion of the network. By decreasing the size of their one-hop neighborhoods we can reduce the infection footprint. Breaking up the super-spreaders takes advantage of the highly heterogeneous topology of real networks~\cite{Pastor-Satorras2015}. 

NodeSplit is an iterative method, which progressively selects the current top-degree node and splits it in two: half of the links remain with the existing node, and half of the links are moved to a newly created node. The total number of peer nodes added equals the number of splits and depends on the available budget.
In practice, the new node can be physical or virtual. A new physical node requires additional hardware, while a virtual node requires computational and memory resources on an existing server. Virtual machines are usually a cheaper and more flexible alternative to adding physical servers to the network. 

Node splitting can be implemented in real networks by using existing traffic management frameworks. For instance,  Microsoft Azure  offers the Traffic Manager service~\cite{TrafficManager}, which allows routing of an application's traffic to different endpoints. Its randomized traffic splitting procedure distributes an application's traffic at random to multiple nodes and can  be used directly to implement NodeSplit.  Network traffic can also be reconfigured to different nodes using SDN-enabled policies~\cite{NetworkViews}.

\vspace{3pt}
\noindent \textbf{Edge Hardening: }
Choosing the optimal edges to secure has been previously investigated in the context of spectral analysis, which refers to the study of the eigenvalues and eigenvectors of the adjacency matrix $A$ of a graph. 
 One particularly important metric in spectral analysis is the largest eigenvalue of the adjacency matrix, $\lambda_1$, because it captures communicability (i.e., path capacity) in a graph. Minimizing $\lambda_1$ was proven to effectively stifle the spread of a virus~\cite{Chakrabarti2008,Prakash2011}.
We investigate the following edge hardening methods:
\begin{enumerate}

\item RandE: Baseline  that randomly chooses $b$ edges from the graph to harden.

\item MET (short for Multiple Eigenvalues Tracking)~\cite{le_2015}, a well-known algorithm that was shown to successfully minimize $\lambda_1$, which utilizes \emph{eigen-scores}~\footnote{The \emph{eigen-score} of an edge $e = (i, j)$ equals to the product of the $i$-th and $j$-th elements of the left and right eigenvectors corresponding to $\lambda_1$, i.e. $u(i) \times v(j)$.} to estimate the effect of removing edges. MET iteratively chooses edges with highest eigen-score to remove until the budget $b$ of edges is reached.

\item NodeSplit + MET hybrid strategy: This combined strategy consists of reconfiguring a small number of nodes with NodeSplit in order to make the edge distribution more homogeneous, followed by edge hardening with MET. We show in this study that this hybrid strategy is able to minimize the leading eigenvalue more than each of the two methods used separately.
\end{enumerate}

\vspace{1pt}
\noindent \textbf{Node Hardening: }
Targeted defense methods in epidemic spreading contain the virus by immunizing a small number of nodes. 
Which nodes to prioritize for immunization is a very relevant question both in social and computer networks, and has been the objective of numerous studies~\cite{BaigAkoglu:WWW2015,Pastor-Satorras2015,torres_2021}. 
From a cybersecurity perspective, immunization  corresponds to  node  hardening  methods such as security patches, system  updates,  and node monitoring  via endpoint agents. The secured nodes become very hard to compromise and are removed from the \texttt{Attacker's Reachability Graph}. 
We analyze four methods for node immunization:

\begin{enumerate}

\item RandN: This is a baseline method, in which the nodes are chosen randomly throughout the network. 

\item Degree: We progressively immunize the most connected nodes, as they contribute highly to the spread of infection.

\item ENS (Effective Network Size): Nodes with the highest effective network size are more likely to act like ``bridges'' between dense clusters~\cite{structural_holes} and monitoring them is likely to prevent attack spreading. 

\item NB (Nonbacktracking Centralities): We explore a recent method~\cite{torres_2021}, which uses the behavior of the spectrum of the nonbacktracking matrix. This method identifies nodes whose removal from the network has the largest impact on the leading eigenvalue of the nonbacktracking matrix.

\end{enumerate}

%\subsection{Isolation}
% Goal: Restrict the attack to smaller portions of the graph. The graph is segmented in multiple connected components (i.e. communities). Since the virus can not jump between components, the attack is contained inside the component where it was initiated.

\vspace{1pt}
\noindent \textbf{Isolation: }
%Communities (also called clusters or modules) are topological partitions of graphs with dense connections within partitions, but sparse connections between different partitions. 
Communities are topological groups of nodes with dense internal connections.
We design and explore community isolation strategies that work in two steps:
First, community detection algorithms are used to identify communities. Many such algorithms are readily available -- we use the well-known Infomap, Leading Eigenvector and the newer Leiden algorithm~\cite{traag2019}. Second, community borders are secured, by hardening either nodes or edges. This translates in securing candidate bridge connections from the \texttt{Attacker's Reachability Graph}, in order to effectively detach the communities of a network and limit the spread of the attack. We study three isolation methods:

\begin{enumerate}
\item CI-Edge, in which all the edges on the borders are secured.

\item CI-Node, in which boundary nodes with highest degree are secured, in decreasing order of their degree. 

%We also explored other methods of ranking boundary nodes to secure, which generally performed worse and are not considered further, including: 1) using betweenness centrality, and 2) using the set cover formulation, that consists in finding the minimal set of boundary nodes that cover (connect to) all edges on the borders, and securing nodes in decreasing order of their coverage.

\item NodeSplit + CI-Edge hybrid strategy. Reconfiguring a small number of nodes before performing community isolation improves the division of a network into modules. This is shown by an increase in modularity~\footnote{Modularity is defined as the number of edges falling within groups minus the expected number in the null model of the network (i.e., an equivalent network with edges placed at random)~\cite{Newman2006}}.
%, as we keep splitting more nodes in this hybrid approach. 

\end{enumerate}

\begin{comment}

system hardening:
https://www.trentonsystems.com/blog/system-hardening-overview

https://www.hysolate.com/blog/system-hardening-guidelines-best-practices/

examples of hardening, lateral movement
https://arxiv.org/pdf/1905.01002.pdf

saha phd thesis:
https://vtechworks.lib.vt.edu/bitstream/handle/10919/82711/Saha_S_D_2016.pdf?isAllowed=y&sequence=1

\end{comment}

\section{Experimental Setup}

\noindent \textbf{Datasets: }
We use an anonymized network flow dataset from an industry partner consisting of 3.4 million nodes and 50 million links. The Critical Watch Report of 2019~\footnote{https://www.newnettechnologies.com/study-finds-majority-of-port-vulnerabilities-are-found-in-three-ports.html} found out that 65\% of vulnerabilities on TCP/UDP ports are linked to only 3 ports: 22, 443 and 80. 
Therefore, we extract and investigate the communication corresponding to a few representative ports: 22 (SSH), 80 (HTTP), 139 (SMB over NetBIOS), 383 (HP data alarm manager), 443 (HTTPS), and 445 (SMB), described in Table~\ref{tab:graphs}. 
For each port, We construct undirected, unweighted, 3-core graphs (i.e.,  the maximal subgraph where all the vertices have degree at least $3$). 
We note the wide range of graph sizes we are investigating, and also the small `Avg Dist' (i.e., the mean vertex-to-vertex distance), which implies that any attack will spread fast within the network.
For other properties illustrated here (Density, Diameter, Transitivity) we refer the reader to~\cite{newman2003structure}.

We discover that our graphs have power-law degree distribution: while the bulk of the nodes have small degree, there is a smaller number of nodes with degrees much higher than the mean value. This property plays a crucial role in devising best defense strategies and it has been shown to be key to explain the success of targeted immunization strategies.

\begin{table*}[!t]
\centering
\begin{tabular}{|c|c|c|c|c|c|c|c|c|}
\hline
Port & Number of & Number of & Mean     & Density  & Diameter  &Avg       & Transitivity   &Transitivity \\ 
     & Nodes & Edges   & Degree    &          &       &Dist  & (global)     & (avg local)    \\ 
\hline
22 &60,825 &333,797 &11 &0.0002 &11 &2.81 &0.0001 &0.181\\
80 &287,156 &1,833,568 &13 &0.00004 &9 &3.01 &0.00003  &0.058\\
139 &1,912 &9,532 &10 &0.005 &9 &2.93 &0.000006 &0.001\\
383 &7,101 &22,910 &6 &0.0009 &7 &3.45 &0.001 &0.236\\
443 &620,096 &4,437,255 &14 &0.00002 &12 &2.80 &0.000002 &0.023\\
445 &317,031 &6,832,418 &43 &0.0001 &10 &2.79 &0.00001 &0.058 \\
\hline
\end{tabular}
\caption{Topological data for the six port-based graphs studied.}
\label{tab:graphs}
\vspace{-25pt}
\end{table*}

\vspace{3pt}
\noindent \textbf{Evaluation measures: }
Let $G$ be the original graph and $G'$ the perturbed graph after applying the defense methods. We use the following evaluation measures, whose importance in quantifying a network's resilience to attacks has been pointed out in previous research~\cite{torres_2021}:

\begin{itemize}
\item\textbf{EigenDrop $\Delta \lambda$ -- drop in the leading eigenvalue: } This metric captures the path capacity reduction within the graph. The leading eigenvalue characterizes the epidemic threshold~\cite{Prakash2011} -- i.e., the regime required for an epidemic to occur.
    Decreasing the leading eigenvalue of the graph essentially increases the epidemic threshold and enables stronger attacks to die out fast. The percentage drop in the leading eigenvalue $\lambda$ is: $\Delta \lambda \% = 100 \times \frac{\lambda - \lambda'}{\lambda}$, where $\lambda'$ is the leading eigenvalue of the perturbed graph $G'$.
    
 \item\textbf{Fragmentation $\sigma$ -- size of the largest connected component relative to the total graph size: }
Let $N$ be the size of the graph, and $N_L$ the size of its largest connected component. We define $\sigma = N_L / N$ as the fraction of nodes contained in the largest connected component. 
The larger the number of nodes that can be reached by the attack, the more damage it can cause. Reducing $\sigma$ enables attack surface reduction by containing the attack within smaller connected components and thus reduces the infection footprint.

\end{itemize}

\section{Evaluation of Network Resilience}
In this section we evaluate how successful are the four types of defenses at increasing the network resilience to SPM.

\begin{figure*}[!t]
\centering
    \includegraphics[width=\linewidth]{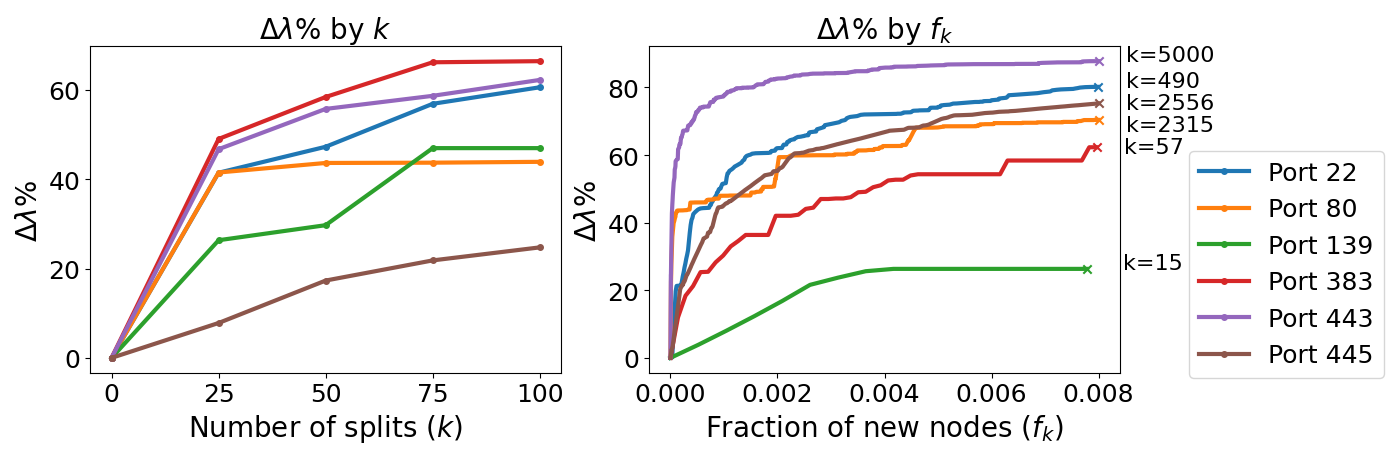}
\caption{NodeSplit: EigenDrop after node splitting, depending on the number of split nodes, $k$ (left), and the fraction of split nodes, $f_k$ (right) (higher is better). A 40-50\% increase is observed on four of the graphs after only 50 splits (left). Steady increase for all ports as the fraction of split nodes increases (right).} 
\label{fig:ld_balancing}
\vspace{-15pt}
\end{figure*}

\vspace{3pt}
\noindent \textbf{Node Splitting:}
The NodeSplit method progressively selects the node with highest degree and transfers half of its edges to a newly created node to balance the number of connections. Thus, the graph becomes more homogeneous in terms of the degree distribution. We analyzed this transition using Alstott et. al~\cite{alstott2013powerlaw}'s mathematical package and report the following findings.
For the smaller graphs (ports 139, 383 and 22), a critical point has been reached before a fraction of 0.2 new nodes were added to the graph; after that, the distribution is closer to exponential, rather than power-law (indicated by the loglikelihood ratio). The trend towards a more homogeneous degree distribution also occurs for the larger graphs, however reaching the critical point requires significantly more splits.

Intuitively, the ``communicability'' in the graph is also decreased, as the fastest spreaders of information, the hubs, have reduced their number of connections. This is captured by a decrease in the leading eigenvalue. Figure~\ref{fig:ld_balancing} illustrates the EigenDrop, $\Delta \lambda \%$, depending on the number of new nodes (which equals the number of splits). On most ports, even a small number of splits leads to a substantial decrease in the leading eigenvalue. For example, just 50 node splitting operations are needed to achieve a 40-50\% lambda drop for graphs whose mean degree is in teens. The long-term trend of a slow but steady EigenDrop increase is visible in Figure~\ref{fig:ld_balancing} (right) for larger graphs, as we keep splitting more nodes. On the other hand, the largest connected component (i.e., our second evaluation metric) is generally preserved.

\begin{comment}
\begin{figure*}[!bt]
\vspace{-15pt}
\centering
  \begin{subfigure}[b]{0.32\linewidth}
    \includegraphics[width=\linewidth]{plots/node_splitting/lambda_drop_100.png}
    \caption{}
    \label{fig:ld_balancinga}
  \end{subfigure}
    \begin{subfigure}[b]{0.32\linewidth}
    \includegraphics[width=\linewidth]{plots/node_splitting/lambda_drop_5000_5.png}
    \caption{}
    \label{fig:ld_balancingb}
 \end{subfigure}
\caption{NodeSplit: EigenDrop after node splitting, depending on the number of split nodes, $k$ (a), and the fraction of split nodes, $f_k$ (b) (higher is better). Large EigenDrop  is achieved  after 25 node splittings (a). Steady EigenDrop increase for all ports as the fraction of splitted nodes increases (b).} 
\label{fig:ld_balancing}
\vspace{-15pt}
\end{figure*}
\end{comment}

\begin{figure*}[!tb]
\centering
    \includegraphics[width=\linewidth]{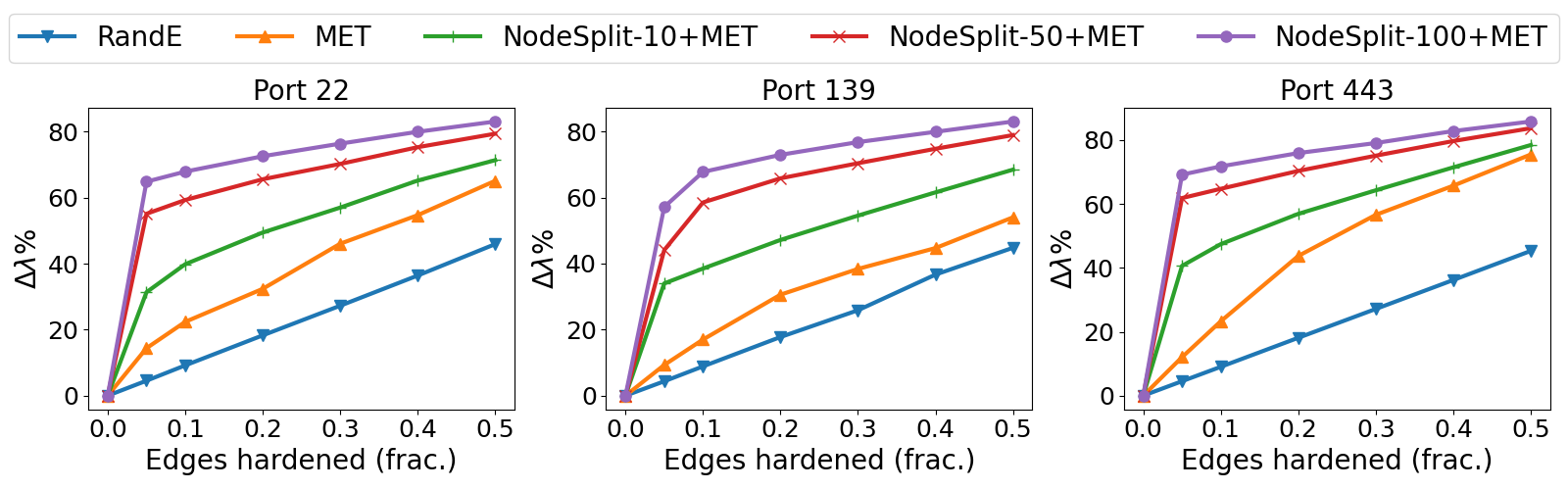}
\caption{Edge Hardening: EigenDrop depending on the percentage of removed edges. NodeSplit-k denotes a number of $k$ splits (higher is better). Hybrid strategies provide additional increase in EigenDrop over MET on all graphs.}
\label{fig:ld_met}
\end{figure*}

\vspace{3pt}
\noindent \textbf{Edge Hardening:}
We compare RandE, MET~\cite{le_2015}, and a hybrid method that combines NodeSplit with MET in Figure~\ref{fig:ld_met}
in terms of EigenDrop. While MET  reduces the leading eigenvalue by itself, the hybrid strategy provides a significant additional drop. For example, at 10\% edges removed (x-axis), the hybrid strategy with 100 split nodes (NodeSplit-100 + MET) almost triples MET's EigenDrop on all graphs. MET is designed to work within a connected graph, and fragmentation (our second resilience metric) is generally negligible.

\begin{figure*}[!t]
\vspace{-10pt}
\centering
    \includegraphics[width=\linewidth]{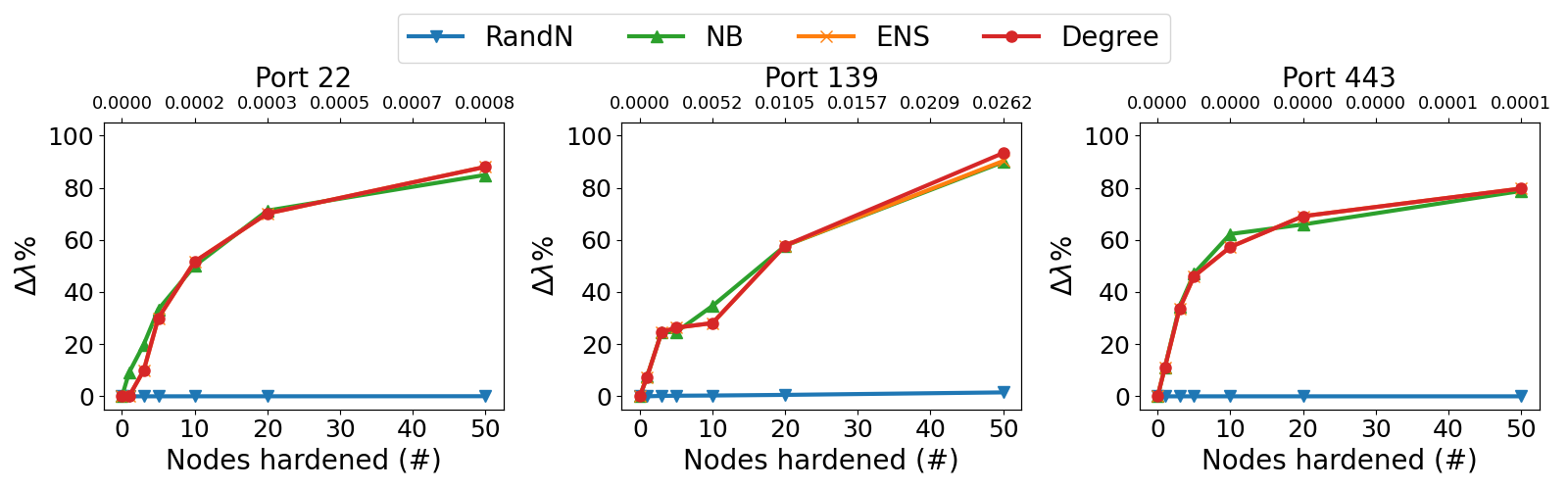}
\caption{Node Hardening: EigenDrop depending on the number of immunized nodes (higher is better). Degree, ENS and NB methods significantly outperform RandN. Top x-axis shows the \emph{fraction} of immunized nodes, which is very small. Similar results on the other port-based graphs.}
\label{fig:lambda_nodes}
\vspace{-15pt}
\end{figure*}

\vspace{3pt}
\noindent \textbf{Node Hardening:}
Next, we evaluate the benefit of ``immunizing'' a small number of nodes to minimize the spread of malware using the following methods: RandN (baseline random node removal), Degree (top-degree nodes), ENS (top nodes according to the effective network size metric), and NB (non-backtracking centralities). 
Figure~\ref{fig:lambda_nodes} illustrates the decrease in the leading eigenvalue after immunizing a number of nodes given by the budget.  
Interestingly, Degree, ENS and NB  deliver very similar performance. This is due to the presence of a highly skewed degree distribution with a few heavily connected star-like nodes which also act as bridges: information needs to go through them to reach other nodes, resulting in large ENS and high path capacity (NB). 
As expected, uniform immunization strategies (RandN) are not effective, as they give the same weight to nodes with high and low connectivity. 

\begin{figure*}[!t]
\centering
\includegraphics[width=\linewidth]{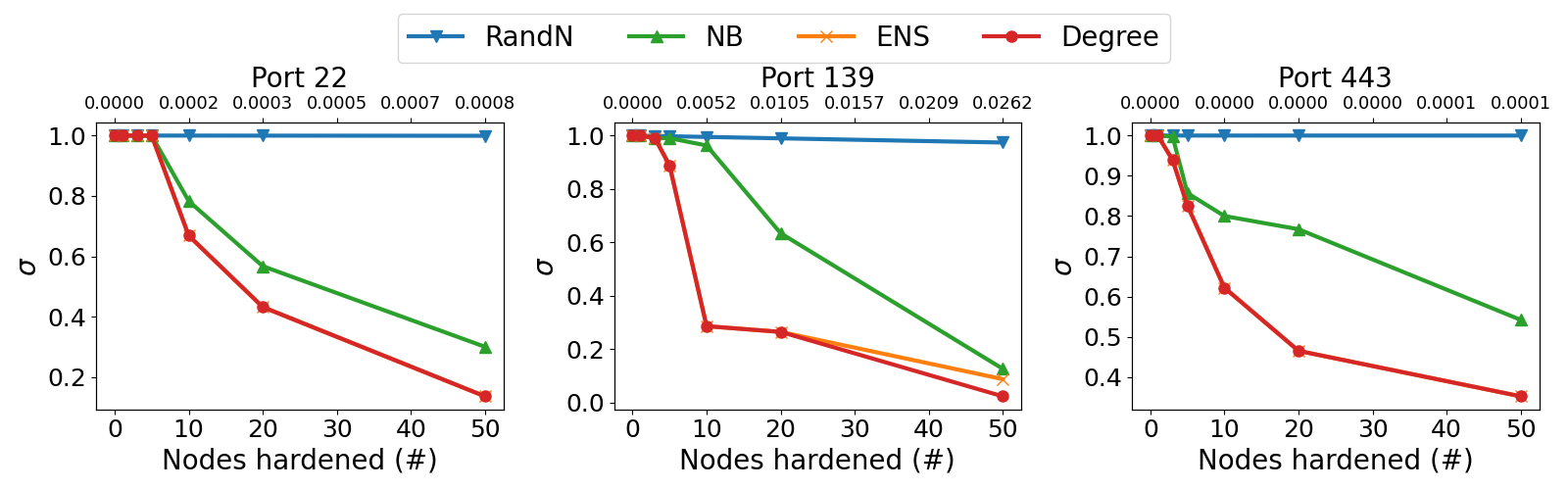}
\caption{Node Hardening: The size of the largest connected component in terms of the original graph size ($\sigma$) (lower is better). Immunizing just 50 nodes on all ports significantly breaks down the largest connected component with all of the methods. Top x-axis shows the \emph{fraction} of immunized nodes.}
\label{fig:fragm_rmnodes}
\vspace{-15pt}
\end{figure*}

How does the graph structure change when the immunized nodes are removed? Figure~\ref{fig:fragm_rmnodes} shows that Degree, ENS and NB exhibit a sharp drop in the size of the largest connected component. In contrast, RandN tends to select low-degree nodes that represent the vast majority of nodes, and whose removal has a low impact on connectivity. On ports 22 and 139, immunizing just 50 nodes with Degree/ENS is enough to break down the largest connected component. Such a rapid disintegration happens due to the highly heterogeneous degree distribution, and was shown to be characteristic of scale-free networks~\cite{Albert00errorand}. 

\vspace{3pt}
\noindent \textbf{Isolation:}
Isolating communities breaks the \texttt{Attacker's Reachability Graph} into smaller connected components and thus decreases the attack footprint. This method uses a community extraction algorithm to identify communities, and then secures the borders. 
We compared three methods for community extraction: Leiden, Infomap and Leading Eigenvector.
Leiden performed better, both in terms of fragmentation achieved and run time, therefore, the experiments in this section use the Leiden method. 

\begin{figure*}[!t]
\centering
  \begin{subfigure}[b]{0.32\linewidth}
    \includegraphics[width=\linewidth]{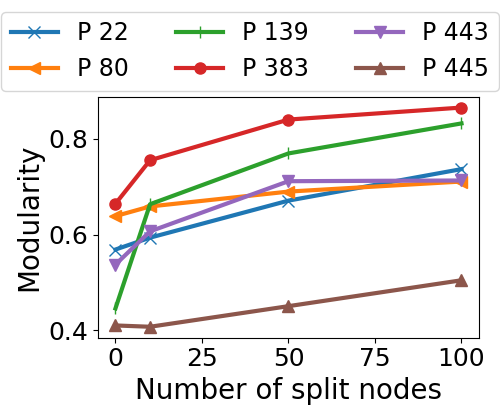}
    \caption{Modularity}
    \label{fig:mod}
  \end{subfigure}
    \begin{subfigure}[b]{0.64\linewidth}
    \includegraphics[width=\linewidth]{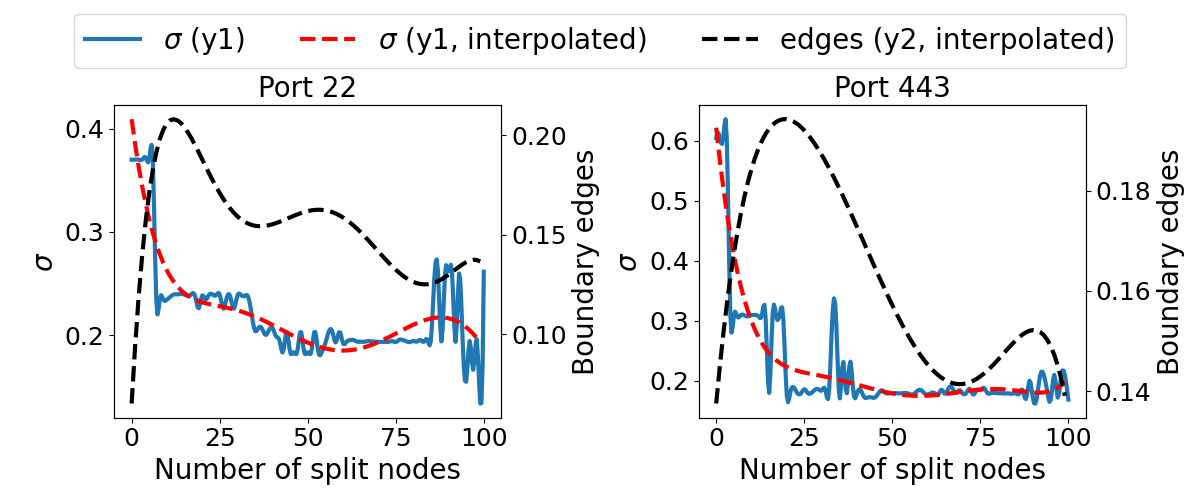}
    \caption{Largest connected component, frac. ($\sigma$)}
    \label{fig:ci1}
  \end{subfigure}
\caption{NodeSplit+CI-Edge hybrid strategy depending on the number of splits. (a): Modularity on all graphs increases. P stands for Port. (b): The size of the largest connected component decreases (y1-axis). At the same time, the boundaries become sparser, with fewer edges (y2-axis) and better isolated communities.}
\label{fig:hybrid_isolation}
\vspace{-15pt}
\end{figure*}

Our experiments revealed that node hardening methods perform similarly at graph level and community level, because many of the most connected nodes are also inter-community ``bridge nodes'' (ports 22, 80, 443, 445).  
However, community isolation offers a viable alternative to immunization when it is preferred to secure edges instead of nodes. Hybrid strategies consisting of NodeSplit + CI-Edge are particularly promising, given that the modularity property of isolated communities increases with the number of splits on all graphs (Figure~\ref{fig:mod}). As communities become more modular, we can achieve a better partitioning of the network. 
This hybrid method is able to reduce the size of the largest connected component significantly. The trade-off between the level of fragmentation obtained and the fraction of boundary edges removed is illustrated in Figure~\ref{fig:ci1}: $\sigma$ decreases, and, eventually, after an initial peak, the communities become more modular, with fewer inter-community edges. 

\section{Evaluation of Infection Spreading}
\label{sec:siidr_sim}
In this section, we investigate which defenses are able to mitigate or even completely eradicate attacks.
To this end, we run stochastic simulations of SIIDR on the communication graphs, using the parameters   estimated from modeling WannaCry from Table~\ref{tab:siidrparams}. ``Patient zero'' $P0$ is a single randomly chosen initially infected node. This is a good approximation, because targeted attacks would be even more promptly stopped by some of the defense methods (i.e., those targeting the super-spreaders, such as node hardening and NodeSplit).
The results are averaged over 500 stochastic instances of ($P0, \beta, \mu, \gamma_1, \gamma_2$) and use the following budgets: 50 nodes for NodeSplit, 50 nodes immunized with Degree, ENS, NB, and a fraction of 0.1 edges removed with MET.

\begin{figure*}[!t]
\centering
  \begin{subfigure}[b]{0.32\linewidth}
    \includegraphics[width=\linewidth]{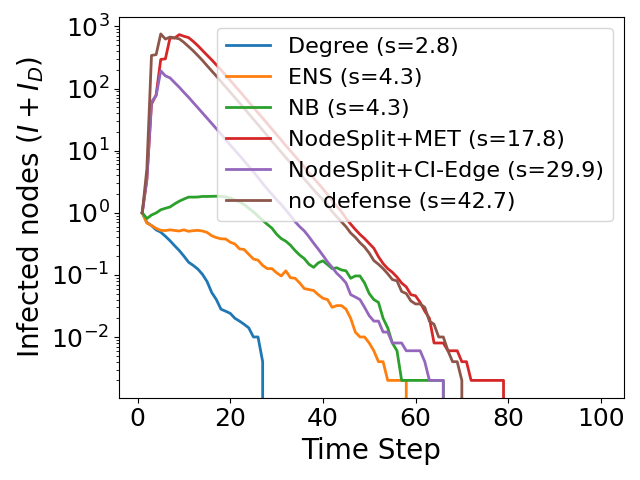}
    \caption{Port 139, wc\_1\_5s}
  \end{subfigure}
\begin{subfigure}[b]{0.32\linewidth}
    \includegraphics[width=\linewidth]{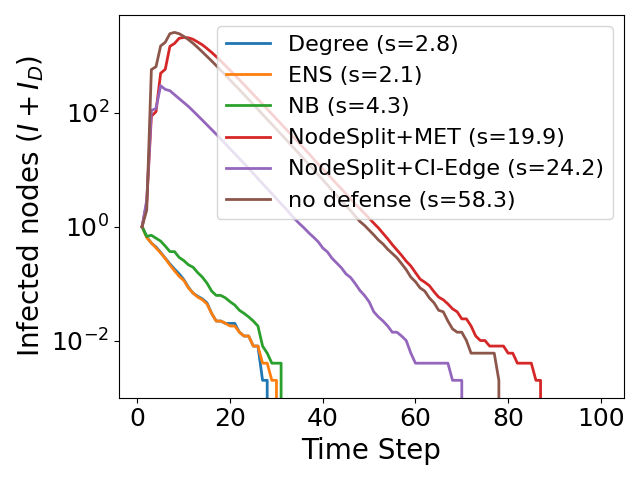}
    \caption{Port 383, wc\_1\_5s}
  \end{subfigure}
    \begin{subfigure}[b]{0.32\linewidth}
    \includegraphics[width=\linewidth]{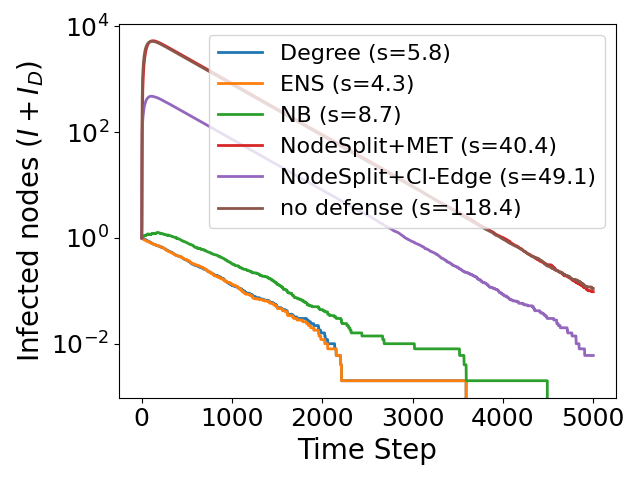}
    \caption{Port 383, wc\_8\_20s}
  \end{subfigure}
\caption{Number of infected nodes ($I + I_D$) in log scale, averaged over 500 simulations, depending on the time step (lower is better). Methods that lead to linear infection decay (i.e., node hardening) are able to prevent outbreaks.}
\label{fig:dieout}
\vspace{-15pt}
\end{figure*}

\vspace{3pt}
\noindent \textbf{Attack Eradication:}
Prakash et al.~\cite{Prakash2011} derived the effective strength  $s = \lambda_1 \times (\beta/\mu)$
for a generic epidemics model that covers SIR and SIS in addition to several other models. 
If $s \leq 1$, then the infection dies out \emph{exponentially} fast, which translates into a linear decay on a log-linear plot~\cite{Chakrabarti2008}.
Our experimental results on ports 139 and 383 presented in Figure~\ref{fig:dieout} show that $s$ is able to capture the ``tipping point’’ for SIIDR as well, with the infection exhibiting a nearly linear decay when $s$ is close to 1.
We observe that, compared to `no defense', the node hardening strategies lead to a $20\times$ decrease in the effective attack strength $s$, while the hybrid NodeSplit-based strategies achieve a $2-3\times$ decrease. For the attack variants studied here, only the node hardening techniques are getting close to a linear infection decay, and, thus, are able to prevent a major outbreak. Hybrid NodeSplit-based methods would be successful in eradicating stealthier attacks (with slower propagation speeds than our attack variants). Attack eradication becomes even more challenging on larger scale-free graphs like ports 22, 80, 443, 445, which are comprised of millions of edges and up to 600K nodes. The difficulty is due to very high $\lambda_1$ values that have been shown to grow with the size of scale-free graphs~\cite{chung2003,Pastor-satorras03epidemicsand}. Implicitly, large $\lambda_1$ results in high effective attack strength $s$, which requires higher budgets of hardened nodes and edges than the ones used here to quickly stop the attack.

\begin{figure*}[!t]
\centering
  \begin{subfigure}[b]{0.32\linewidth}
    \includegraphics[width=\linewidth]{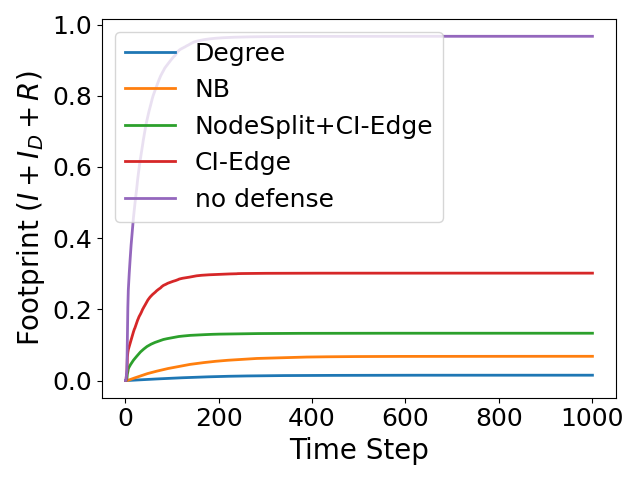}
    \caption{Port 22, wc\_1\_1s}
  \end{subfigure}
\begin{subfigure}[b]{0.32\linewidth}
    \includegraphics[width=\linewidth]{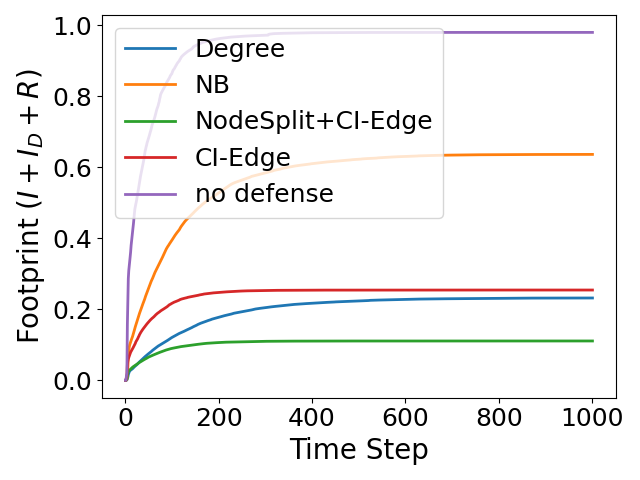}
    \caption{Port 80, wc\_1\_10s}
  \end{subfigure}
    \begin{subfigure}[b]{0.32\linewidth}
    \includegraphics[width=\linewidth]{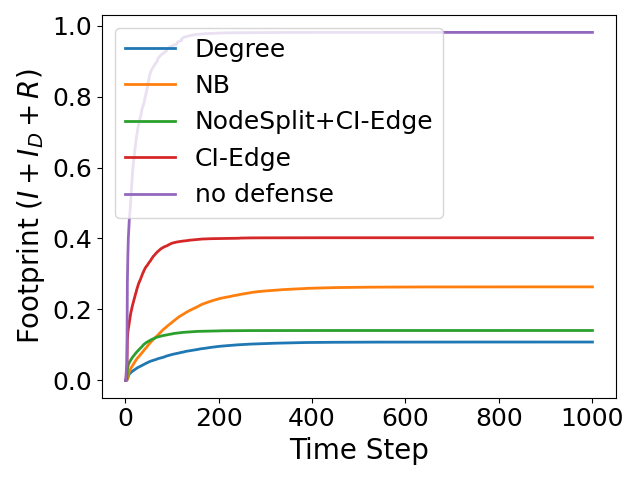}
    \caption{Port 443, wc\_4\_5s}
  \end{subfigure}
\caption{Infection footprint ($I+I_D+R$) depending on the time step (lower is better). Degree and ENS overlap, hence, ENS was omitted. Significant reduction of the infection footprint was obtained on all ports across all attack variants after applying the defenses (other port/variant scenarios omitted due to space limit).
}
\label{fig:footprint}
\vspace{-15pt}
\end{figure*}

\vspace{3pt}
\noindent \textbf{Attack Mitigation:}
However, even at low budgets, our defenses can still achieve major improvements in terms of minimizing the infection footprint, defined as the total fraction of nodes in compartments $I$, $I_D$ and $R$. We will demonstrate these results next, using the larger port-based graphs. Defense methods that reduce the largest connected component, such as node hardening and community isolation strategies, are the most successful at attack mitigation, because they contain the attack within smaller segments of the network. Figure~\ref{fig:footprint} shows a reduction of the infection footprint from about 97\% with `no defense' down to 1\% on port 22 (with Degree), 11\% on port 80 (with hybrid NodeSplit+CI-Edge), and 11\% on port 443 (with Degree) for the attack variants illustrated here. Note that these are strong attacks, that lead to almost all nodes being infected when no defenses are employed. We obtained substantial attack mitigation across the board, for all ports and variants. 

\vspace{3pt}
\noindent \textbf{Defense Generality:} In this section, we presented  defense results against self-propagating malware implemented with the SIIDR model, using parameters estimated for the 7  WannaCry malware variants from Table~\ref{tab:siidrparams}. We have additionally run SPM attacks with a range of parameters for infection and recovery rates in SIIDR, varying the ratio $\beta / \mu \in (0,1]$, and we observed similar findings. Typically, node hardening strategies, such as Degree and ENS, achieve the lowest infection footprint. Node splitting and community isolation defenses are also effective, particularly in hybrid strategies. While we selected the SIIDR model for our experiments here (given its lowest AIC scores on our malware variants), we believe that our defenses will be similarly effective against other epidemiological models for malware propagation, including SIR and SIS.

\section{Related Work}
Network robustness had been studied across multiple domains like infrastructure, communication and social networks, in order to understand the functioning and vulnerabilities of complex interconnected systems. Key metrics for measuring robustness (including the spectral radius and largest connected component used in this study) have been proposed and used in several papers~\cite{Albert00errorand,BaigAkoglu:WWW2015,chencascadian2018,freitas2022graph,tong_2010}.
A large body of work has looked at modeling the spread of epidemiological processes in networks~\cite{Chakrabarti2008,Pastor-satorras03epidemicsand,Pastor-Satorras2015,Prakash2011}. Building up on this, several strategies that propose to stop the infection by manipulating the adjacency matrix of the graph have been developed, including edge removal algorithms~\cite{le_2015,tong_2012} and node removal techniques that use centrality measures such as degree centrality, betweenness centrality, PageRank,  eigenvector centrality, Katz centrality, X-centrality, etc.~\cite{freitas2022graph,tong_2010,torres_2021}. 

Rey~\cite{rey2015} provides a comprehensive review of compartmental models for malware modeling. Mathematical proposals for the modeling and evaluation of malware dynamics include~\cite{guillen2019security,mishra2014dynamic,toutonji2012stability,zhu2012modeling}. Other works focus on malware propagation in specific settings, like wireless sensor networks~\cite{ojha2021improved} or VMs  under the infrastructure as a service (IaaS) architecture~\cite{gan2020dynamical}. Closer to our work on malware fitting, Levy et al.~\cite{levy2020modeling} use the classical SIR model to identify the rate of infection and other parameters from real traces. 

Reconfiguring communication generally has tackled the reverse problem compared to our NodeSplit method: rather than preventing attack propagation, the goal of previous research was to rewire edges in order to have alternative paths if a hub fails~\cite{chan2016_rewire,louzada2013}. Community isolation has been studied from the point of view of the attacker, using module-based attacks to fragment social, infrastructure and biological networks~\cite{RequiaoDaCunha2015}.

\section{Discussion and Conclusions}
Recent large-scale cyber attacks such as WannaCry and Mirai have demonstrated how pervasive the risk of self-propagating malware has become. With cyber threats looming, it is important to proactively address vulnerabilities in networks to minimize the impact of an attack. From our extensive experiments with real-world graphs and realistic modeling of  WannaCry attacks, we have gained several cybersecurity insights. We summarize these insights in a set of recommendations for security teams to increase network resilience against SPM attacks.

First, we note that SPM malware relies on network connections to spread. Therefore, the best defenses must create topological structures that prevent SPM from spreading without impeding the transmission of legitimate traffic. Closing unused ports, enforcing firewall blocking rules, and creating access control policies for communication flows are some of the possible edge hardening techniques that help reduce the attack surface. Second, the best defenses exploit the inherent hierarchy in networks. Attacks that target hubs or bridges have the potential to be the most devastating. The most effective defense is to identify key super-spreaders and allocate  security budgets to protect them. Third, reconfiguring communications via access control policies can greatly increase the robustness of the network against attacks. We have shown in our experiments that it is efficient to split the most connected nodes in half to balance their connections; we expect high gains even with smaller budgets. 

Finally, focusing on communities within the network makes it possible to prioritize security measures under limited budgets. A modular approach can also leverage densely connected partitions by isolating them from the rest of the network to prevent a large-scale infection spread. Moreover, it addresses the issue of vanishingly small epidemic thresholds of large scale-free networks, making it easier to stop a self-propagating attack before it becomes an epidemic.

In our experiments, we analyzed in detail the impact of these recommendations as a function of their cost in terms of topological changes. From a cybersecurity perspective, hardening nodes through security updates and patches is the most straightforward method and can help prevent major SPM attacks, including the WannaCry malware studied here. Monitoring nodes through intrusion-detection systems is more costly, especially when traffic analysis is performed over a large number of connections, but it can prevent other cyber attacks. Node splitting is based on managing network traffic in a cyber network, which can be implemented with existing traffic management solutions (available in cloud data centers) or SDN-enabled policies. These solutions provide performance improvements in addition to security, but they induce additional implementation costs.

\section*{Acknowledgements} This research was sponsored by PricewaterhouseCoopers LLP and the U.S.~Army Combat Capabilities Development Command Army Research Laboratory (DEVCOM ARL) under Cooperative Agreement Number W911NF-13-2-0045. The views and conclusions contained in this document are those of the authors and should not be interpreted as representing the official policies, either expressed or implied, of DEVCOM ARL or the U.S.~Government. The U.S.~Government is authorized to reproduce and distribute reprints for Government purposes notwithstanding any copyright notation here on.

\bibliographystyle{splncs04}
\bibliography{references}

\end{document}